\documentclass[sigplan,screen,nonacm]{acmart}

\usepackage{soul}
\usepackage{xcolor}

\AtBeginDocument{%
  \providecommand\BibTeX{{%
    \normalfont B\kern-0.5em{\scshape i\kern-0.25em b}\kern-0.8em\TeX}}}

\setcopyright{acmcopyright}
\copyrightyear{2022}
\acmYear{2022}
\acmDOI{XXXXXXX.XXXXXXX}

\acmConference[Conference acronym 'XX]{Make sure to enter the correct
  conference title from your rights confirmation emai}{June 03--05,
  2018}{Woodstock, NY}
\acmPrice{}
\acmISBN{}

\begin{document}

\title{Self-Sovereign Identity in a World of Authentication: Architecture and Domain Usecases}


\author{Morgan Reece}
\affiliation{%
 \institution{Mississippi State University}
  \country{Mississippi State, MS, USA}
}\email{mlr687@msstate.edu}
 
 \author{Sudip Mittal}
\affiliation{%
  \institution{Mississippi State University}
  \country{Mississippi State, MS, USA}
  }
\email{mittal@cse.msstate.edu}

\renewcommand{\shortauthors}{Reece, et al.}

\begin{abstract}
Self-Sovereign Identity (SSI) is projected to become part of every person's life in some form. The ability to verify and authenticate that an individual is the actual person they are purported to be along with securing the personal attributes could have wide spread implications when engaging with third party organizations. Utilizing blockchains and other decentralized technologies, SSI is a growing area of research. The aspect of securing personal information within a decentralized structure has possible benefits to the public and private sectors. 

In this paper, we describe the SSI framework architecture as well as possible use cases across domains like healthcare, finance, retail, and government. 
The paper also contrasts SSI and its decentralized architecture with the current widely adopted model of Public Key Infrastructure (PKI).

\end{abstract}



\maketitle

\section{Introduction}
The ability to purposefully verify your identity in the digital realm is the foundation for most, if not all, online activities. This online identity is something that we protect at varying levels depending on the negative impact of a compromise on our lives. With a larger portion of our lives being transacted online, people are getting more savvy in the way they manage their digital identities. It used to be common practice to use the same `simple' password for all your accounts. Now that is seen as a recipe for disaster, knowing that if one online account is leaked, then all your accounts are potentially exposed \cite{Beck2016}. Organizations are helping users by moving them toward safer online identity practices requiring multi-factor authentication and stronger passwords. These steps have helped improve online identity security immensely, but the architecture of the standard online identity management system is inherently insecure. The development of blockchain and decentralized identities has enabled the creation of the Self-Sovereign Identity (SSI) framework. The SSI framework has huge possibilities of implementations across many domains where there is a drive toward individual privacy \cite{Tykn2022}.

The development of these frameworks aims to improve the overall security of a user's Personal Identifiable Information (PII). Examples of PII include a user's name, address, passport number, driver’s license number, taxpayer identification number, patient identification number, financial records, etc. Personal Health Information (PHI) is one of the most sought after PII by hackers for the primary reason that identities used in healthcare are tied to Social Security Numbers (SSN) which never change for an individual \cite{Accoutnable2022}. Other PHI, such as home address, and phone number are used by organizations to determine identity can change. 
Impersonation using stolen PHI puts the individual in a scenario which can lead to severe financial loss. SSI usage in the healthcare industry could help secure user data and limit the proliferation of patient PHI. Healthcare providers that utilize an SSI model would only have access to patient PHI with the approval/permission from the patient.


Similar to Public Key Infrastructure (PKI), SSI uses asymmetric keys as the encryption architecture as well as digital signatures to ensure secure communication between peers \cite{Sorokin2021}. 
PKI utilizes a group of \textit{central} authorities to store the association database of identities, certificates, and public encryption keys. Conversely, SSI employs Decentralized Ledger Technology (DLT) to associate Decentralized Identifiers (DID) and public encryption keys. In an SSI implementation, DID and the public portion of their symmetric key pair are the only publicly accessible information associated with an entity. The entity retains their identity information details which they share as they see fit.


This paper describes the SSI architecture and its major components, along with a comparison between the mechanics used in PKI and SSI. We also examine how SSI can be utilized in different domains like healthcare, finance, retail, and government, where identity authentication and data security are expected and in some cases required by regulation.

The architecture presented in this paper is an identity framework utilizing blockchain, Decentralized Identity (DID), as well as other decentralized and cryptographic technologies. The framework is focused on the idea of \textit{self-sovereignty} of one's identity. Self-sovereignty can be defined as having control over the specifics of your own identity. This includes the user's ability to be able to share specific identity attributes and acquired credentials depending on the intended recipient of the information. The user will present a profile to the recipient, where the profile includes only those identity details required by the recipient for a particular operation.


The main goals of the SSI architecture is to fulfilling the CIA triad; Confidentiality, Integrity, and Availability. Other frameworks fulfill these goals through different architectural implementations relying on different structures and technologies to meet these goals. The presented framework prioritizes information security and user control of access.


The overall impact and proliferation of SSI through different domains has not been fully realized. Most domains have some level of need for associating identity with a collection of data or transactions. SSI has the ability to deliver personal information protection and control. Many systems built on the SSI framework are being developed to meet these different needs with the promise of higher level of security and greater control afforded to the user.



The rest of the paper is organized as follows - Section \ref{relwork} presents some background and related works. We describe the SSI architecture and its components in Section \ref{arch}. SSI is compared with PKI in Section \ref{comp}. Section \ref{use} covers a number of development efforts utilizing SSI that are taking place across multiple domains. We discuss the impact of SSI on the identity industry in Section \ref{discuss}. We conclude in Section \ref{conc}.

\section{Related Work}\label{relwork}

In this section, we present some related work and background on Identity, Distributed Ledger Technologies and Blockchains, Authentication, and Trust.

\subsection{Identity before Blockchains}

Public Key Infrastructure (PKI) has been developed over the last 35 years, built utilizing centuries of research in cryptography \cite{stapleton2012concise}. Many people consider PKI to be the defacto solution for managing entity identities on the internet. PKI is used to encrypt communication as well as establishing identity of users and systems. 
The history of the concept of Self-Sovereign Identity (SSI) can be traced back to 1991 and the Pretty Good Privacy (PGP) project \cite{OpenPGP2016}. The idea of a Web-of-Trust (WoT) was created as a PGP project \cite{Stahl2018}. In WoT there are a few entities/other users trusted by a particular user. 
As this user continues to add trusted entities to their WoT, other users/entities trust them as well. These other users have a collection of trusted entities/users that is different from the first user. In this model there is a relationship established (e.g. primary, secondary, tertiary, ...) between all users of the PGP project and results in the creation of a full WoT. The development of WoT was a monumental step in decentralized authentication.

In 1996, Carl Ellison presented the idea of identity without certification authorities \cite{Ellison1996}. Ellison presented the idea of binding the certificate of the issuer in a manner that is recognized by the verifier. The idea of not having a central authority to be the issuer and verifier was an innovative concept that was years ahead of its time. The decentralized aspect of the authorization is clearly seen in the current SSI architecture.

Microsoft proposed a federated ID model when they created MS Passport \cite{MSPowerU2016}. The concept was for a single login to allow access to multiple systems. A good idea to improve user experience and ease access to multiple systems/sites. However, the design relied upon Microsoft as the central authority to manage authentication and authorization. This model was no better, and in some ways worse than a traditional central authority.

\subsection{Decentralized Identity, Distributed Ledger Technologies, and Blockchains}

The next big development in decentralized identity was Identity Commons \cite{allen2016} and the 
creation of the Internet Identity Workshop (IIW) working group in 2005. The working group put forward the idea of user-centric identity. There is a clear connection between these ideas and the decentralized architecture in the SSI framework. The IIW has long supported OpenID which an individual could utilize as a psuedo-SSI. The main issue is that to truly mimic SSI using OpenID, a user would need advanced technical knowledge to implement it properly.

Nakamoto wrote a paper in 2008 describing the bitcoin blockchain \cite{Nakamoto2008}. The decentralized nature of blockchain along with the addition of technological advancements has allowed for the development of SSI. Leveraging the creation of blockchain along with the WoT and IIW, the Decentralized Identity Foundation (DIF) was formed in 2017 \cite{DIF2021}. The DIF is a consortium of technology innovators in SSI. The member of DIF include, but not limited to Microsoft, Hyperledger (Linux Foundation), Accenture, and Sovrin. Hyperledger and Sovrin are both non-profit organizations that rely on contributing member to work collaboratively to define and development SSI frameworks and best practices. Sovrin utilizes the Hyperledger Indy codebase for it system \cite{Tobin2017}.

The Hyperledger Indy framework is being used in a number of development/research efforts. The open source aspect of these framework makes them a great platform for researchers to investigate, propose, and develop new innovations in the SSI framework. Being part of Hyperledger or Sovrin enables organizations to help drive development as well as provide technical guidance with regard to the direction that the projects should follow \cite{Tobin2017}.

There are businesses that are developing what could be considered SSI, but because of the centralized nature of the company which controls the authorization of the users, they cannot truly be considered SSI. One such implementation is 1Kosmos which utilizes a digital wallet on your smart phone to manage PII and credentials \cite{1kosmos2022},


\begin{figure*}[ht]
    \centering
    \includegraphics[scale=.95]{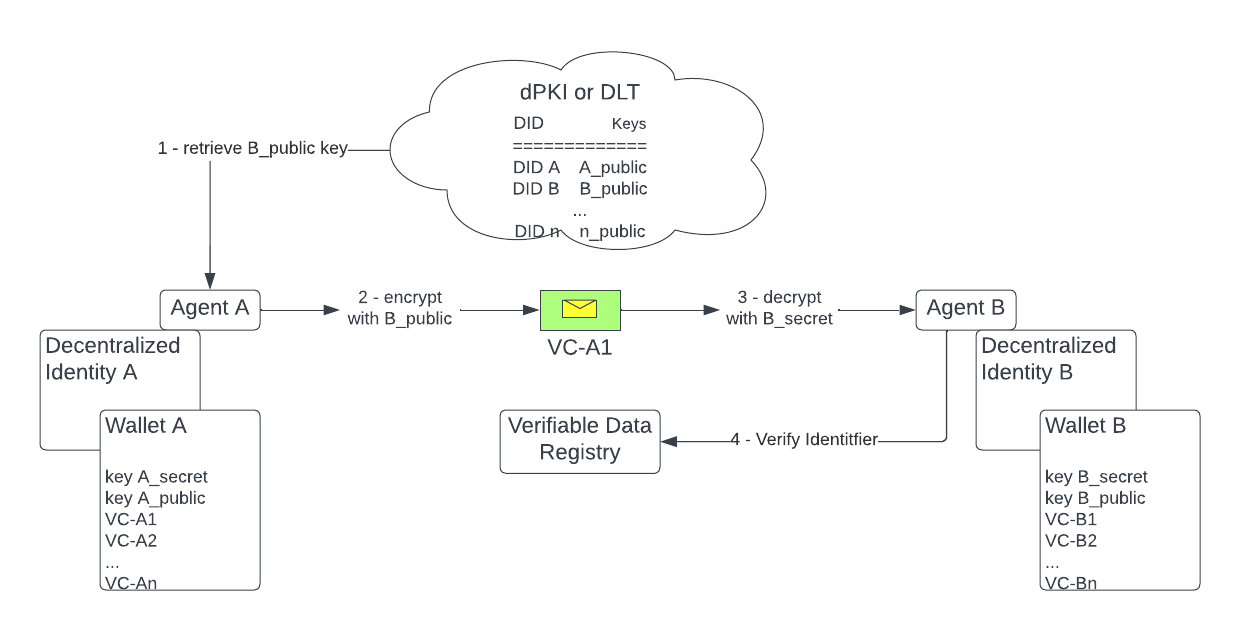}
    \caption{\textbf{SSI Architecture Flow} Here, Agent B is verifying Agent A. Agent A, acting on behalf of an entity, and utilizing the data in Wallet A, (1) key B\_public is retrieved by Agent A and (2) used to encrypt and send a message, a verifiable credential (VC-A1) to Agent B. The message is (3) received by Agent B and decrypted by its private key B\_secret. Entity B then (4) verifies the credential VC-A1 against the Verifiable Data Registry.} 
    \label{fig:Arch.png}
\end{figure*}

\subsection{Authentication}

The concept of authentication is closely related to identity. An entity can claim an identity \cite{BlueCheck2021}; however, until that claim is authenticated, it is just a claim and therefore access to protect objects (systems, data) should not be granted. Passwords have traditionally been used in digital systems to authenticate a user's identity. Passwords are characterized as knowledge based identity authentication. PINs, passwords, and personal knowledge answers all fit into the knowledge based category. Two other type of authentication categories exist; `What you have?', and `Who you are?'. `What you have?' relates to a physical object of which you have possession. This can be a mobile phone, memory/smart card, or another certificate bearing object. `Who you are?' relates to biometric information. This information can be fingerprint, iris scan, or facial scan. Utilizing more than one of these categories of authentication is known as Multi-Factor Authentication (MFA). With the advent software that is able `crack' passwords, MFA has become a defacto standard by which modern systems are secured, especially healthcare and financial systems. The ability prove identity to access secure systems is critical to maintaining confidentiality as part of an overall security program. 

\subsection{Trust}

Within an SSI architecture, there is a level of trust that is required. The trust within the peer-to-peer relationship that is used for communication and identity verification is developed by utilizing other peer relationships to confirm the veracity of an entity's claim \cite{Stahl2018}. The Web of Trust (WoT) project presents a model of trust based on reputation. Utilizing nodes within the WoT allows for trust to be extended to parties that may not have a direct relationship with an entity. The challenge in the WoT are nodes that try to manipulate the trust network to allow more trust to be offered to an entity that should not be trusted. In a peer-to-peer relationship, extending trust must be guarded.

\section{Self-Sovereign Identity (SSI) Architecture}\label{arch}


The foundation of SSI is the decentralized identifiers architecture presented by W3.org \cite{w3c2022}. Core components include Decentralized Identifiers (DID) v1.0, architecture, data model, and representations. The task is to present credentials to another entity for verification/authentication. Figure \ref{fig:Arch.png} shows a simple communication between two DIDs sending a verifiable credential. {Here, Agent B is the verifier and Agent A needs to be verified.} Agent A, acting on behalf of entity A, and utilizing the data in Wallet A, (1) public key B\_public is retrieved by A and (2) used to encrypt a message, a verifiable credential (VC-A1) to B. The message is (3) received by B and decrypted by its private key B\_secret. Entity B then (4) verifies the credential VC-A1 against the Verifiable Data Registry. 

The SSI architecture is composed of seven key technologies defined by the W3C. The seven technologies are:

\begin{itemize}
    \item Decentralized Identifiers
    \item Verifiable Credentials
    \item Decentralized Public Key Infrastructure
    \item Blockchain and Distributed Ledger Technology
    \item Verifiable Data Registry
    \item Agents
    \item Digital Wallets
\end{itemize}


Next week describe each of these in detail.

\subsection{Decentralized Identifiers (DID)}
Decentralized Identifiers (DID) are globally unique identifiers structured similar to a universally unique identifier (UUID) but with modifications to enable cryptographic identity/security \cite{w3c2022}. The first difference is there is no centralized authority managing the identities. DID's use a decentralized mechanism like a Decentralized Ledger Technology (DLT) \cite{Frankenfield2021}. The other difference is that the DID has cryptographic properties. Key pairs are generated as part of the DID address and can be used to prove ownership/identity in the same fashion as digital signatures. The DID Auth protocol is a challenge-response process and therefore has the ability to replace the username/password structure of a centralized authentication system \cite{Gataca2021}. The immutability of the blockchain and the use of cryptogrpahic proofs to affirm ownership makes the DID platform a rich environment for the development of authoritative identity verification.

\begin{figure}[hb]
    \centering
    \includegraphics[scale=0.75]{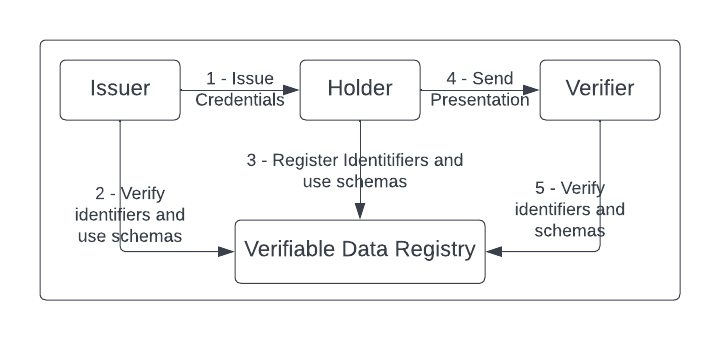}
    \caption{\textbf{Issue/Holder/Registry} Within the process of issuing verifiable credentials, the Issuer (1) issues credentials to the Holder which is the entity being authenticated using the credential. The Holder (3) submits the credentials and its use schema to the Verifiable Data Registry, where the Issuer (2) verifies the identities and use schema of the credentials. The Holder (4) also send the credentials in a presentation to the Verifier, where the Verifier (5) send the identities to the Verifiable Data Registry where the identity and schema are verified.}
    \label{fig:VC-flow.png}
\end{figure}

\subsection{Verifiable Credentials}
Verifiable credentials (VC) are exactly what one would assume them to be, credentials that are able to be verified by the issuers of the credential. There are three main parts of the VC; the issuer, the holder, and the registry (See Figure \ref{fig:VC-flow.png}). The issuer defines the credential scheme to the registry as well as issuing a specific credential to the holder. The holder will register the credential with the registry. To complete the transaction of verification, a third party verifier will receive a credential presentation from the holder, and then confirm authenticity of the credential with the registry \cite{Bolgouras2022}. The credential presentation is a `view' of the credential that is defined as specific attributes of the credential that the verifier requires to be seen to confirm authenticity of the credential. The benefit of this limited view is that the presentation does not need to disclose all the information in the credential, but only what is agreed upon by the holder and verifier as to what information is `enough' to confirm the veracity of the credential.

\subsection{Decentralized Public Key Infrastructure (DPKI)}
Decentralized Public Key Infrastructure (DPKI) is a part of the overall decentralized model of identity verification \cite{Singh2017}. While the DPKI is not directly related to the individual identity verification, it is a significant security system in which the SSI systems rely upon to provide system/URI identity verification. In the event that the PKI is compromised, then the SSI environment is able to be breached, and therefore bring into question the veracity of the whole system. DPKI resolves multiple vulnerabilities of current Public Key Infrastructure Certificate Authority (PKI/CA) systems. DPKI utilizes blockchain and smart contracts to provide system identity confirmation.

\subsection{Blockchain and Distributed Ledger Technology (DLT)}
What is commonly known as blockchain is a Distributed Ledger Technology (DLT). Blockchain gained notoriety with the popularization of Bitcoin. Bitcoin was an early implementation of cryptocurrency based on a 2008 whitepaper written by Satoshi Nakamoto \cite{Nakamoto2008}. Cryptocurrencies utilize the trust and integrity of the blockchain to manage/document the transactions between parties. Because of the distributed nature of blockchain, availability is also provided in the architecture. Confidentiality is not a concern for cryptocurrencies which rely on the ability to see all transactions. SSI utilizes the all three aspects of the security triad; Confidentiality, Integrity, and Availability. 
Blockchain can be implemented in four configurations, where there are two settings for two parameters. The four configurations are as follows:
\begin{itemize}
    \item Public - Permissionless
    \item Public - Permissioned
    \item Private - Permissionless
    \item Private - Permissioned
\end{itemize}

While cryptocurrencies are usually Public - Permissionless where they can be seen and added to by anyone, SSI implementations are typically Public - Permissioned which can be seen by anyone, but one needs special permission to add to the blockchain. \cite{Tobin2017} The Private - Permissioned is able to support the integrity desired in an SSI. Each block in the blockchain contains hashed transactions creating an immutable block as part of the distributed ledger. 

\subsection{Verifiable Data Registry}
A Verifiable Data Registry manages the life-cycle and verification of credentials. The database system that acts as a verifiable data registry can take multiple forms driver's licenses, passports, and other government ID databases are examples of verifiable data registries \cite{Dunphy2018}. 

\subsection{Agents}
Agents are software applications that run either in the cloud or on an `edge' device such as a mobile device or browser \cite{Reed2016}. The agents acts on behalf of an entity to access digital wallets as well as performing other cryptographic operations. The benefit of the agents is the increased privacy provided by using the agent.

\subsection{Digital Wallets}

\begin{figure}[hb]
    \centering
    \includegraphics[width=8cm]{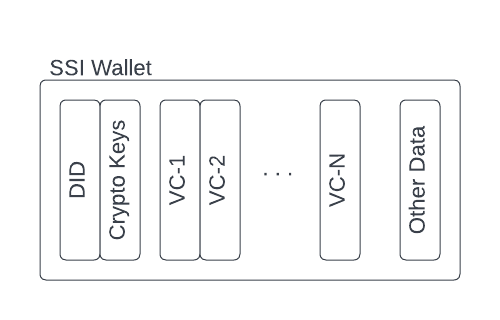}
    \caption{\textbf{Digital Wallet} Information stored within an SSI Wallet. Items stored include DID, crypto-keys, multiple verifiable credentials, and other data.}
    \label{fig:Wallet.png}
\end{figure}

Digital Wallets can be software or hardware implementation of a data store. Digital wallets work in concert with agents to perform multiple functions related to the management of cryptographic information such as digital signatures, key pairs life-cycle management. The digital wallet can take many forms including being supported as a Software as a Service  (SaaS) function \cite{Soltani2021}. SSI Wallets can contain multiple verified credentials as well a other data structure. Figure \ref{fig:Wallet.png} shows the primary data structures utilized in an SSI Wallet. Items stored include DID, crypto-keys, multiple verifiable credentials, and other data. The other data mentioned can be Non-Fungible Tokens (NFTs), specific cryptocurrencies, or other personal data.

\section{Comparison with Public Key Infrastructure (PKI)}\label{comp}

\begin{figure}[ht]
    \centering
    \includegraphics[width=8cm]{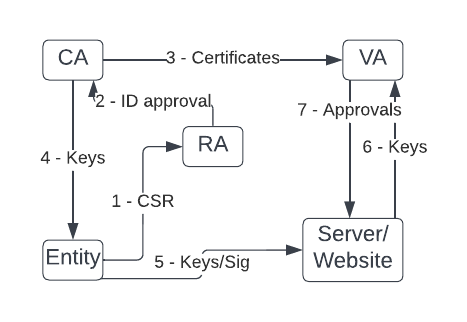}
    \caption{\textbf{PKI Architecture} (1) Entity submits CSR (Certificate Signing Request) to RA (Registration Authority). (2) RA sends ID approval to CA (Certificate Authority). (3) CA sends certificates to VA (Verification Authority). (4) CA sends keys back to the requesting entity. (5) Keys/Signature are sent to or installed on a server or website. (6) Keys are sent to the VA for verification. (7) Verification/approval is returned confirming the authenticity of the keys/signature.}
    \label{fig:PKI.png}
\end{figure}


Public Key Infrastructure (PKI) can be characterized as a central store of identity registration where cryptographic keys are associated with a registered identity. Identities can be users or devices. The inclusion of a PKI certificate is used to confirm the authorized identity associated with the certificate. There is a hierarchy of trust that is established within the PKI services that allow for the life-cycle management of the certificates as well as the authenticity verification of the certificates.

PKI utilizes a challenge-response process to verify the identity authenticity of the certificate holder (See Figure \ref{fig:PKI.png}). This is similar to the DID authentication process which is also a challenge-response loop.

A key \textit{difference} between SSI and PKI is the decentralization of SSI. The decentralization of SSI allows for exclusive peer-to-peer coordination of transactions \cite{Singh2017}. These transactions can include identity verification as well as exchange of information or digital asset. For PKI, there is an exclusive function performed that relates to cryptography. The certificate and key pair that are issued as a result of registration with a PKI Certificate Authority (CA) can be used for encryption as well as identity verification function as in digital signatures. SSI has the ability within its architecture to support any type of digital asset as well as tokens such as NFTs. The construct of a smart contract within the SSI architecture allows for enumerable implementations supporting peer relationship agreements within the blockchain. PKI certificates/cryptogrpahy are used almost exclusively for encryption and identity verification.

Another key difference between SSI and PKI is the hierarchical nature of PKI as opposed to the flat peer-to-peer nature of SSI. Within the PKI, there is a structure of hierarchy that has the root Certificate Authority (CA) at the top of the structure. {Root CAs maintain the database of certificates and share it with subordinate CAs, which are the systems that respond to queries}. The Registration Authority (RA) receives requests for new certificates as well as performing a number of other administrative tasks. They also offload the life-cycle management of certificates from the CAs.

The central database of a PKI makes the CA's a significant target for attackers who wish to leverage vulnerabilities in the security of the PKI for their advantage. Compromising a CA would result in an attackers ability to direct users to malicious websites masquerading as actual websites. The decentralized nature of SSI as well as the use of blockchain as the data store creates a immutable data structure that distributed across the internet. To compromise this architecture, an attacker would have to be able to compromise multiple systems simultaneously which high unlikely if not completely impossible. 

\section{Usecases}\label{use}
While SSI is still considered in it's infancy, there are a number of development efforts that are taking place across multiple domains. The immutability and security of the SSI architecture make it a compelling framework for many functions where users wish to move to a fully digital environment. In this section, we describe some usecases possible in healthcare, finance, retail, and government.

\subsection{Healthcare}

The creation and adoption of the Health Insurance Portability and Accountability Act (HIPAA) \cite{HIPAA2003} and Health Information Technology for Economic and Clinical Health (HITECH) Act \cite{HITECH2017} instituted guidelines on the protection and portability of Personal Healthcare Information (PHI). These laws outline how healthcare providers must manage patient information. Business associates are companies and individuals who utilize, manage, or engage in business with healthcare organizations and access, store, process, or control patient information. Utilizing SSI to store PHI would enable patients to have their medical history immediately available when seeing a new doctor or visiting an emergency room. There is also the aspect of increased security afforded when using SSI as opposed to a central database. SSI also allows a patient to control which personal and health information is shared with a healthcare provider or payer. 

\begin{figure}[htp]
    \centering
    \includegraphics[scale=0.7]{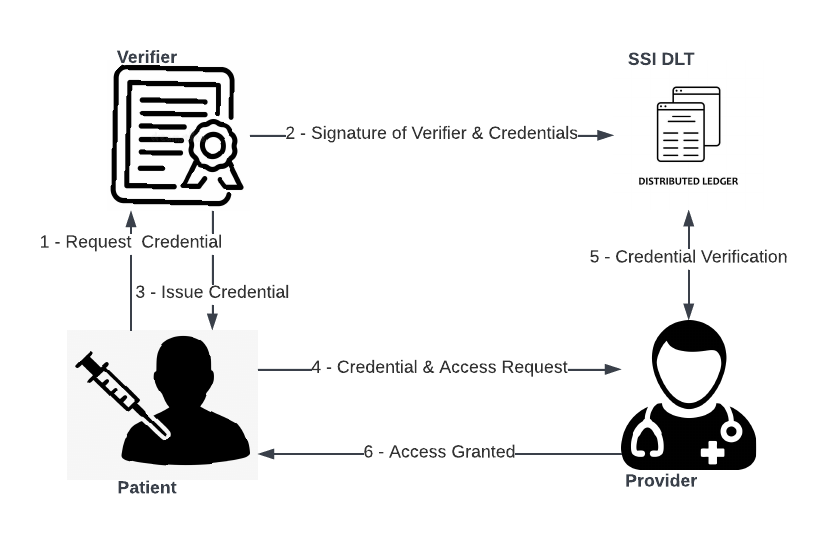}
    \caption{\textbf{Healthcare SSI Architecture} (1) Patient requests a credential (2) Verifier send credential and signature to DLT (3) Verifier issues credential to patient (4) Patient sends request to Provider with Patient credentials, requesting access (5) Provider sends credentials to DLT for verification where DLT responds with approval (6) Provider replies with access granted.} 
    \label{fig:Healthcare.png}
\end{figure}

The authentication of patient identity/credential would follow the steps in figure \ref{fig:Healthcare.png}. (1) Patient requests a credential from an authority able to issue the required credential. (2) Verifier send credential and its own signature to a DLT to be used as a point of verification for the issued credentials. (3) Verifier issues credential to patient where the patient stores them in their digial wallet. (4) Patient sends request to Provider, with Patient credential issued from Verifier, requesting access to Patient data. (5) Provider sends credentials to DLT for verification where DLT responds with approval. (6) Provider replies with access granted.

\subsection{Finance}

The ability for banks and other financial institutions to leverage SSI could bring about significant improvements to the customer experience \cite{Keil2022}. Utilizing SSI, banks would be able to streamline the audit process as well as the ability to onboard new customers. Couple this improved user experience with the improved security and lower back-office costs, the motivation to incorporate SSI into the banking industry has tangible and quantifiable benefits.

\subsection{Retail}
There has been significant development in the area of digital currency that impact the retail industry. The idea of making expenditures as an individual without the insertion of a middle-man to manage the transfer of funds has been widely accepted. However, SSI goes beyond just the ability to make a peer-to-peer transaction that is safe, secure, and private \cite{Tobin2017}. SSI being incorporated into the retail/e-commerce ecosystem could have a significant impact on several aspects of these financial transaction. A few benefits are the increased security of the financial transaction, the reduction in fraudulent purchases, and the lack of need to store customer payment information. This last benefit would reduce the number of lost/stolen customer payment information, therefore improving the purchasing experience for the customer as well as the retailer.

\subsection{Government}
In our modern societies we rely on the governments to provide individuals with credentials that affirm our claim to our personal identity. In the USA, prospective employees must submit government approved documentation to demonstrate one's identity as well as nationality. A few of the documents that are acceptable to be used for identity proofing are a state issued driver's license, social security card, and a passport. These forms of credentials are difficult, but not impossible to forge. The use of SSI to maintain a digital wallet with verified credentials from a government agency would provide a significant improvement to the identity proofing process as well as securing specific PII that would not need to be held in the government database. India has a national digital identification system which issues a residents an Aadhaar number \cite{India2019}. To receive an Aadhaar number, a person must provide the following demographic and biometric information:

\begin{itemize}
    \item Name
    \item Date of Birth (verified) or Age (declared)
    \item Gender
    \item Address
    \item Mobile Number (optional) and Email ID (optional)
    \item Ten Fingerprints
    \item Two Iris Scans
    \item Facial Photograph
\end{itemize}

With this information being collected and technology being developed to help secure personal information, there have been a number of proposals to coalesce multiple government identities housed on one card. A proposal in 2019 suggested that this one card could hold voter registration, bank card, passport, and Aadhaar card information \cite{Misra2019}.

A significant challenge perceived in the use of SSI are the vulnerabilities that can be found in the system components. As some of these SSI components like digital wallets store all consolidated PII, an implementation software bug or an exploitable system vulnerability, would result in significant personal and financial concern to the user. 

\section{Discussion}\label{discuss}
As is the case in the development of many technologies, the concepts for SSI have long been established before the foundational technologies were developed to efficiently and effectively implement the innovative ideas. SSI has been gaining market exposure over the last 5 years including consortiums being established 10 years ago but finally releasing alpha level solutions/products in the past 5 years. Many corporations are looking at SSI to deliver passwordless authentication solutions. Security professionals and attackers have known for years that users are the biggest security vulnerability. This is why phishing emails remain one of the top attack vectors every year. One of the main exploitation methods within phishing emails is an embedded link to a fake login page. A user entering in their username/password is handing their credentials over to attacker that they can later use to login into company resources and conduct a malicious attack. If the company had setup passwordless/SSI authentication, then there would be no credentials to steal and therefore the phishing attack would fail to exploit the user. 

While the SSI model presents the idea that an individual controls their credentials and attributes, that is not always 100\% true. When credentials are verified, there is an issuer that has the ability to revoke/expire these credentials. Therefore, there is a decentralized ledger and entity controlled wallet that manages the presentation of credentials, there are central authorities that are able to revoke credentials.

\section{Conclusions}\label{conc}
After almost 30 years since the creation of the concepts that form the foundation of SSI, the identity industry is investing significantly in development of frameworks to support its use across multiple domains. The development of blockchain and decentralized identity create a synergy in technology that is able to support the notion that an individual should have the ability to control how their personal information is shared with other people, organizations, and companies. The ability to verify an individual's claim of identity and their personal credentials without the need to reference a central authority creates a network of trust that can return the power of identity back into the hands of the individual.

In this paper, we have presented a SSI architecture which will support multiple domain requirements as well as the similarities and differences to the current PKI architecture being used across the internet today. We also discussed specific benefits and usecases possible in healthcare, finance, retail, and government. We elaborate on how these industries could revolutionized the customer/user experience in authentication.


\bibliographystyle{unsrt}
\bibliography{export.bib}


\end{document}